\documentstyle[epsfig,11pt]{article}
\def\bc{\begin{center}}
\def\ec{\end{center}}
\def\beq{\begin{equation}}
\def\eeq{\end{equation}}
\def\noi{\noindent}
\def\pp{p_{\perp}}

\title{\bf Jet energy loss due to multiple  scattering
in the nucleus }
\author{M.Braun \\
Department of high-energy physics,\\
University of S. Petersburg, 198904 S. Petersburg, Russia }
 \date{July 2009}

\begin{document}
\maketitle
\medskip
\noi{\bf Abstract}

Energy loss of a jet due to multiple collisions in the nuclear
target is studied in the framework of the perturbative QCD with an
infrared cutoff. An iterational procedure for its calculation is
developed, which allows to reliably find up to 16 successive
collisions in rescattering. The calculated shift of the scaling
variable reaches values of the order 0.1 at medium $x$ and
transverse momentum $p$ in the interval $5-10$ GeV/c. The shift is
found to be independent of the atomic number of the target and
energy.

\section{Introduction}
Jet quenching in nuclear matter due to gluon emission has long been
a subject of detailed studies, related to the observed behavior
of particle spectra in heavy-ion collisions at RHIC ~\cite{cassol}.
 Much attention
has also been devoted to a simpler quenching mechanism due to hard elastic
collisions ("collisional quenching"), which  is formally of the leading
order in $\alpha_s$ as compared to gluon emission corresponding
to inelastic hard collisions. It is expected that quenching due to
elastic collisions is proportional to the length $L$ travelled by the
jet inside the nucleus, whereas gluon emission seems to lead to
quenching proportional to $L^2$ ~\cite{BDMP}. However this conclusion has been derived
for superhigh energies. The relative strength the two sources of
quenching at present energies is not clear both from the theoretical
and experimental points of view ~\cite{armesto}.

Most calculations of the collisional energy loss (CEL)
 refer to the propagation of a fast parton in the homogeneous
quark -gluon plasma ~\cite{pla1,pla2,pla3,pla4}.
The study of parton energy loss in
a realistic nucleus is a harder task. It requires using the
multiple scattering formalism beyond the leading Glauber approximation
with full attention to the interplay between the longitudinal and transverse
momenta. First calculations made in
~\cite{cata} indicated that elastic collisions diminish the jet energy
quite considerably, up to 40\% at LHC energies. However due to technical
difficulties the authors of ~\cite{cata} were able to push their
calculations of multiple hard scattering of the jet inside the nucleus
only up to the  first three collisions. This is  quite insufficient even for
RHIC energies at which, as our calculation show, at least 6 collisions are to
be taken into account. At the LHC energies this number grows to 12.
To obtain reasonable results for
the overall quenching due to elastic collisions one has to search
for methods which allow to study  multiple
parton collisions with the loss of  energy with a reasonable accuracy.
This paper presents an
iterative approach which seems to give satisfactory results at both
the RHIC and RHIC  energies.

Our framework follows that of ~\cite{catre, actre,bratre} which considers
multiple scattering of a parton existing inside the fast projectile
long before the actual collision. The scattering itself is due to the
QCD perturbative interaction in the lowest order, with a relatively
high transferred transverse momentum and a loss of energy determined
by the kinematics. Naturally we introduce an infrared cutoff at low
transferred momenta (by an effective gluon mass) and our results depend
on the value of this cutoff. On physical grounds we choose it to lie in
the region of 1.5$\div$ 2.0 GeV/c. With such values our iterative procedure
gives reliable results up to 16 collisions at the LHC energy.

From the start we have to stress  that we consider only passage of
a jet through the nucleus. The study of its hadronization is postponed for
future publications. We study the cross-sections as a function of the transverse
momentum $p$ of the observed jet and its scaling variable $x$.
For the study of collisional energy loss of a jet due to rescattering
in the nucleus use of these variables is natural.
Hadronization, necessary to study the observed
parton spectra, can  be conveniently calculated in these variables.
To pass to c.m.rapidities instead of $x$ one can use the
standard relation
$ y=\ln (xW/p)$
where $W$ is the c.m. energy for proton-proton collisions.

The CEL effect was
found to be quite noticeable. For values of $p$ in the interval
$5-10$ GeV/c and $x$ of order 0.5 the decrease in the scaling variable was
calculated to lie in the interval $0.09-1.14$. However a curious result is that this
decrease does not depend on the energy nor on the atomic number of the target.
The latter implies that CEL does not depend on the length of
the rescattering path inside the nucleus.
Cross-sections themselves exhibit more or less the expected behaviour: they fall
with the inclusion of CEL. This fall is only weakly energy dependent,
so that the general picture at the RHIC and LHC energies remains practically the same.
As a curiosity at the RHIC energy we found that the standard nuclear suppression factor
$R_A$ at small values of $x$ actually grows with the introduction of CEL.
This abnormal behaviour however is not seen at higher values of $x$ nor at the LHC energies.


\section{ Glauber-like parton rescattering.}

To formulate our model and have a benchmark for the following calculations
we briefly reproduce here the derivation of the inclusive jet production
in hadron-nucleus or nucleus-nucleus scattering in the pure Glauber approximation
without loss of energy. The details may be found in the original derivation
in ~\cite{bratre}.

We assume the
nuclear state $|A>$ to be a superposition of states with a different
number $n$ of partons, each characterized by its scaling variable $x$
and impact parameter $b$ combined into argument $z=\{x,b\}$:
\begin{equation}
\mid A>=\sum_{n}\int \prod_{i=1}^n d^3z_{i}\Psi_{A,n}(z_{i})\mid n,z_{i}>.
\end{equation}
If the observed parton originates from the A-nucleus (moving along the
z-axis), then the wave function  of the latter is for a given $n$:
\begin{equation}
\Psi_{A,n}(z_{1},z_{2},...,z_{n})=\sqrt n
\psi(z_{1})_{\alpha}\tilde\Psi_{A, n-1}(z_{2},...,z_{n}).
\end{equation}
Here $\psi_{\alpha}(z)$ denotes the wave function of the observed
parton, and $\alpha = \{x,p\} $, where $x$ and $p$ - are the scaling
variable and transverse momentum respectively.

The inclusive cross-section of  parton production at a given
impact parameter $\beta$ for the  nucleus-nucleus (AB) collision is given by
the following expression
\[
\frac{d \sigma}{d^2 \beta}= \sum _{nl}n \int dz_1dz_1^ \prime
\psi_\alpha (z^\prime_1)\psi^*_\alpha(z_1)d\tau_A(n-1)d\tau_B(l)
\Psi^*_{A,n-1}(z_1^ \prime,z_2 ,:,z_n)\]\beq \times
\Psi_{A,n-1}(z_1,:,z_n)| \Psi_{B,l}(u_j)|^2
[S^*_{nl}(z^\prime_1,:,z_n|u_1,:,u_l)-1]
[S_{nl}(z_1,:,z_n|u_1,:,u_l)-1],
\end{equation}
where $d\tau(n)$ is the pase volume for the state with $n$ partons.
This expression contains contributions of both hard and soft
parton interactions. The latter interactions are disregarded
because they cannot be studied in the
perturbative QCD. The contribution of hard collisions is given by
the product $S^*_{nl}\times S_{nl}$.

Further transformations rely on the assumption of factorization of
nuclear S-matrix into a product of elementary S-matrices
corresponding to parton-parton interactions, which is the essence of
the Glauber model:
\begin{equation}
S_{nl}(z_{1},...,z_{n}\mid
u_{1},...,u_{l})=\prod_{i=1}^n\prod_{j=1}^l s_{ij}.
\end{equation}
Here $s_{ij}$ is S-matrix for the interaction of parton $i$ of the A
nucleus with parton $j$ of the B nucleus, while $u_{j}$ - refers to
parton variables of nucleus B. We assume factorization of
parton distributions:
\begin{equation}
|\Psi_{A(B),l}(u_j)|^2=\frac{1}{l!}e^{-\langle
l\rangle}\prod_{j=1}^l\Gamma (u_j),
\end{equation}
with $\langle l\rangle=\int d^3u\Gamma_B(u)$ and $\Gamma_B(u)$
denoting the product of the parton distribution inside the nucleon
and nuclear profile function,
\begin{equation}
\Gamma_{A(B)}(u)=T_{A(B)}(c)P_{A(B)}(\omega ).
\end{equation}
As a result, we get the final expression for the inclusive
cross-section of jet production in AB collisions with hard rescattering taken into
account:
\begin{equation}
\label{eq1} (2\pi )^2\frac{d\sigma_{AB}}{d^2\beta dxd^2p}=\int
d^2bd^2re^{ipr}T_A(b-\beta )P_A(x)e^{-T_B(b)F_B(x,0)}\{
e^{T_B(b)F_B(x,r)}- 1\},
\end{equation}
where
\begin{equation}
\label{eq2}
 F_B(x,b)=\int \frac{d^2q}{(2\pi)^2}\int d\omega P_B(\omega)
I(x,\omega,q)e^{ipb}
\end{equation}
and $I(x,\omega,q)$ is the distribution in the transverse momentum $q$
in the
elastic collision of two partons with their
scaling variables $x$ and
$\omega$.

For hadron-nucleus collisions one has to change $T$ into $\delta$-function
and integrate over $\beta$ to obtain
\begin{equation}
\label{eq11}
(2\pi )^2\frac{d\sigma_{hA}}{dxd^2p}=\int
d^2bd^2re^{ipr}P(x)e^{-T_A(b)F_A(x,0)}\Big\{
e^{T_A(b)F_A(x,r)}- 1\Big\},
\end{equation}
where $P(x)$ is the parton distribution in the incoming hadron.
\section{Energy loss in a single elastic parton collision}
\subsection{Kinematics}

Following the picture of multiple parton scattering let an
incoming (massless) parton (4-momentum $p=(p_+,p_-,p_{\perp})$)
scatter on a target parton
(4-momentum $k=(0,k_-,0)$). The final momenta are
$p'=(p'_+,p'_-, \pp-q)$ and $k'=(k'_+,k'_-,q)$ where $q$ is the
transverse part of the transferred momentum. The scaling variables
of the initial and final projectile parton are $x$ and $x'$, those of
the target parton $w$ and $w'$. The parton collision c.m. energy squared
is $sxw$ where $s$ is the overall c.m. energy squared.
The longitudinal momenta conservation gives
\beq
p'_++\frac{q^2}{2k'_-}=p_+,\ \ \frac{(\pp-q)^2}{2p'_+}+k'_-=
\frac{\pp^2}{2p_+}+k_-.
\label{conserv}
\eeq
The energy loss of the incoming jet is
determined by the difference
\beq
\Delta p_+=p_+-p'_+=\frac{q^2}{2k'_-},
\eeq
which transforms into the corresponding decrease of the
jet scaling variable $x$.

From Eqs. (1) we conclude
\beq
2k'_-p'_++q^2=2k'_-p_+,\ \ 2k'_-p'_++(p_{\perp}-q)^2=2p'_+
\left(\frac{p_{\perp}^2}{2p_+}+k_-\right)
\eeq
and thus
\beq
q^2-2k'_-p_+=(p_{\perp}-q)^2-2p'_+
\left(\frac{p_{\perp}^2}{2p_+}+k_-\right).
\eeq
This allows to express $k'_-$ via $p'_+$:
\beq
k'_-=\frac{1}{2p_+}\Big[2p'_+(\frac{p_{\perp}^2}{2p_+}+k_-)+
q^2-(p_{\perp}-q)^2\Big]
\eeq
and obtain an equation:
\beq
\left(\frac{p'_+}{p_+}-1\right)
\Big(2p'_+(\frac{p_{\perp}^2}{2p_+}+k_-)+
q^2-(p_{\perp}-q)^2\Big)+q^2=0.
\label{eq}
\eeq
This is
a quadratic equation from which one determines the initial projectile
longitudinal momentum $p_+$ in terms
of its final longitudinal momentum and other variables describing
the collision: $w$, $q$ and $p_{\perp}$.
Denoting
\beq
x=\frac{x'}{1-u}
\label{defu}
\eeq
one finds from (\ref{eq}
\beq
u=\frac{1}{p_{\perp}^2}\Big\{p'_+k_-+{\bf p}_{\perp}{\bf q}-
\sqrt{(p'_+k_-+{\bf p}_{\perp}{\bf q})^2-
p_{\perp}^2q^2}\Big\}.
\eeq

For small transferred momenta $u$ is proportional to
$q^2$  and then
\beq
u=\frac{q^2}{swx'},
\label{defuu}
\eeq
so that
\beq
x=\frac{x'}{1-q^2/(swx')}
\label{xxprime}
\eeq
and is independent of $p_{\perp}$, although the latter may be
comparatively large as a result of accumulation of several transferred
momenta.
We use this simplified formula in our
calculations.

Note that condition $0<x<1$ leads to the restriction $q^2<sw(x'-{x'}^2)$,
which limits the region of integration in $q^2$, $x'$ and $w$.
From this inequality it also follows that in any case
$q^2<s(x'-{x'}^2)\leq s/4$.

\subsection{Parton-parton cross-section}
In principle one should take into account partons of different flavor both
in the projectile hadron and target nucleus. Although it is straightforward,
it considerably enhances the calculation time. To economize on the latter,
following \cite{eskola}, we considered the effective gluon jet  defined by the
density
\beq
P(x,Q^2)=g(x,Q^2)+\frac{4}{9}\Big(q(x,Q^2)+\bar{q}(x,Q^2)\Big),
\label{defp}
\eeq
where $g$,$q$ and $\bar{q}$ are the gluon, quark and antiquark densities respectively.
Correspondingly for the parton interaction we took the gluon-gluon one with the cross-section,
integrated over the
target gluon distribution  ~\cite{eskola}
\beq
I(x',q)\equiv\frac{(2\pi)^2x'd\sigma}{dx'd^2q}=
(2\pi)^2\frac{9\alpha_s^2}{2q^4}
\int\frac{dw'}{w'}[xG(x)]_{x'+q^2/sw'} [wG(w)]_{w'+q^2/sx}
\Big(1-\frac{q^2}{sxw}\Big)^3,
\eeq
Here, according to our notation, $x$ and $w$ refer to initial gluons
and $x'$ and $w'$ refer to final gluons.
We pass to the integration over the initial scaling variable of the
target
\beq
I(x',q)=
(2\pi)^2\frac{9\alpha_s^2}{2q^4}
\int\frac{dw}{w-q^2/sx'}[xG(x)]_{x'+q^2/sw'} wG(w)
\Big(1-\frac{q^2}{sxw}\Big)^3.
\eeq
From (\ref{defu}) and (\ref{defuu}) one finds
\[ w-q^2/sx'=w(1-u),\ \ 1-\frac{q^2}{sxw}=1-u+u^2,\]
so that the cross-section becomes
\[
I(x',q)=
(2\pi)^2\frac{9\alpha_s^2}{2q^4}
\int\frac{dw}{w}xG(x) wG(w)
\frac{(1-a+a^2)^3}{1-a}\]\beq\simeq
(2\pi)^2\frac{9\alpha_s^2}{2q^4}
\int\frac{dw}{w}xG(x) wG(w)
(1-u)^2
\label{crsec}
\eeq
where the second approximate form takes into account that
$q^2$ and consequently $u$ are assumed to be small.
The cross section $I(x',w,q)$ entering our formulas for the $n$-fold
cross-section is obtained from  this one by dropping the
partonic distributions and integration over $w$.

\section{Loss of energy}
Eqs. (\ref{eq1}) or (\ref{eq11}) have an obvious physical interpretation.
Developing the exponents in powers of $F$ one gets, say, for hA collisions
the expression for the $n$-fold rescattering
\beq
(2\pi )^2\frac{d\sigma_{hA}^{(n)}}{dxd^2p}=\frac{1}{n!}\int
d^2bd^2re^{ipr}P(x)T^n_A(b)
\Big\{
\Big(F_A(x,r)-F_A(0,r)\Big)^n- F_A^n(0,r)\Big\}
\label{nfold}
\end{equation}
At fixed $r$ one finds the product of $n$ inclusive cross-sections
$F_A(x,r)$ summed with similar products in which some of the
inclusive cross-sections are substituted by the total cross-sections
$F_A(x,0)$ with a minus sign. The second term in (\ref{nfold})
eliminates the product which does not contain inclusive cross-sections
at all. In the momentum space the product goes over into a convolution.
So physically this term corresponds to a number $0<k\leq n$ of
consecutive inclusive cross-sections and  $(n-k)$ total cross-sections
distributed among these inclusive cross-sections in all possible ways.

This indicates how one can take into account  the loss of energy
during rescattering.  Each inclusive cross-section in our approximation
corresponds to the elastic scattering with a change of transverse momentum.
If after $n$ such collisions  the observed scaling variable of the jet is
$x$ then before the $n$-th elastic collision and so after the
$(n-1)$-th collision  the jet has  its scaling variable $x_{n-1}>x$.
Similarly before the $(n-1)$-th collision and after the $(n-2)$-th collision
the jet has its scaling variable $x_{n-2}>x_{n-1}$ and so on. In the first
elastic collision the scaling variables before and after the collision
are $x_0$ and $x_1$ respectively. If the loss of energy would be
independent of the scaling variable of the target parton $w$ then
this loss would be accounted for by a simple
substitution
\beq
F_A^n(x,r)\to\prod_{i=1}^nF(x_i,r)
\label{xtoxi}
\eeq
where $x_i$ are the scaling variables of the jet after the $i$-th
elastic collision with $x_n=x$ the scaling variables of the observed jet.
However, as we have seen, the change in $x$ in fact depends on $w$, so that
this substitution has to be done under the sign of the integration over all
scaling variables $w_i$ of the target partons.

Another problem is to distribute the total cross-sections
among the elastic ones with correct scaling variables.
Obviously the total cross-sections
appearing between the $k$-th and $(k+1)$-th elasic ones are to depend
on $x_k$. For instance for $n=2$, with the loss of energy independent
of $w$,
the original Glauber expression
\beq
F^2(x,r)-2F(x,r)F(0,r)
\eeq
should be substituted by
\beq
F(x_1,r)F(x_2,r)-F(x_2,r)F(x_1,0)-F(x_2,r)F(x_2,0),\ \ x_2=x
\eeq
However the $w$-dependence again requires making this substitution
inside the integral over all $w$'s.

A convenient way to obtain constructive formulas which take into account the
loss of energy is to set up an iteration procedure allowing to find
the $(n+1)$-fold rescattering contribution in terms of the $n$-fold one.

\subsection{Iteration procedure}
It is more convenient to work directly in the momentum space.
Consider a sequence of $n$ collisions, some elastic and some
total. Call the contribution of the amplitude which ends with
the $n$-th elastic collision, with nuclear factor separated,
$X_n(x,p)$ where $x$ and $p$ are the scaling variable and
transverse momentum of the jet.
Then the total contribution with any number of elastic
collisions after the last inelastic one is
\beq
Y^{tot}_n(x,p)=\sum_{k=0}^nX_{n-k}(x,p)[-\sigma(x)]^k,
\label{tdef}
\eeq
where $\sigma(x)$ is the total cross-section of the projectile parton
and by definition
\beq
 X_0(x,p)=(2\pi)^2\delta^2(p).
 \eeq
From
(\ref{tdef}) a relation follows
\beq
Y^{tot}_n(x,r)=X_n(x,r)-\sigma(x)Y^{tot}_{n-1}(x,r).
\label{defimprop}
 \eeq
 Note
that $Y^{tot}_n(x,r)$ includes an improper term corresponding to the case
when there are only elastic collisions:
\beq
T^{tot}_n(x,p)=Y_n(x,p)+[-\sigma(x)]^n(2\pi)^2\delta^2(p).
\eeq
Here $Y_n$ is a contribution from $n$ collisions with at
least one inelastic. One easily finds that also
\beq
Y_n(x,p)=X_n(x,r)-\sigma(x)Y_{n-1}(x,p).
\label{defprop}
\eeq

Our basic equation connects $X_n$ with $Y^{tot}_{n-1}$. Under the sign
of integration over the target scaling variable $w$ we have just to
convolute $Y^{tot}_{n_1}(x_{n-1},p)$ with $F(x_n,p)$ to obtain
\beq
X_n(x,p)=\int \frac{d^2q}{(2\pi)^2}dw_nP(w)
I(q,x,w)Y^{tot}_{n-1}(x_1,p-q).
\label{iter0}
\eeq
Here
$x$ is the scaling variable of the projectile AFTER the $n$-th
collision, $q$ and
$w$ are the transferred momentum and the scaling variable of the target
in this last inelastic
collision. The scaling variable $x_1$ is the one BEFORE the $n$-th
inelastic collision considered as a function of collision variables
\beq
x_1=f(q,x,w).
\eeq
It is generally greater than $x$, which corresponds to
collisional quenching. The explicit expression of function $f$
is given by (\ref{xxprime}). The momentum $p-q$
is the projectile momentum BEFORE the last inelastic collision.
With only one elastic collision
\beq
X_1(x,p)=)=\int \frac{d^2q}{(2\pi)^2}dwP(w)I(q,x,w)
(2\pi)^2\delta^2(q-p),
\eeq
which corresponds to
\beq
Y^{tot}_0(x,p)=(2\pi)^2\delta^2(p)
\eeq
and follows from our assumption that the projectile has no transverse
momentum before all collisions.

Separating the improper contribution to $Y^{tot}_{n-1}$ in (\ref{iter0})
we get
\beq
X_{n}(x,p)=\int \frac{d^2q}{(2\pi)^2}dwP(w)
I(q,x,w)\Big\{Y_{n-1}(x_1,p-q)+\Big(-\sigma(x_1)\Big)^{n-1}
(2\pi)^2\delta^2(p-q)\Big\}
\label{iter1}
\eeq
and using relation (\ref{defprop}) get the final form of the iteration
equation
\[
Y_{n}(x,p)=\int \frac{d^2q}{(2\pi)^2}dwP(w)
I(q,x,w)\Big\{Y_{n-1}(x_1,p-q)+\]\beq
\Big(-\sigma(x_1)\Big)^{n-1}
(2\pi)^2\delta^2(p-q)\Big\}-\sigma(x)Y_{n-1}(x,p)
\label{iter2}
\eeq

One starts iterations with
\beq
Y_0(x,p)=0
\eeq
and successively determines $T_1$, $T_2$ and so on.
At fixed $b$ the final contribution to the
inclusive cross-section is obtained as
\beq
\frac{(2\pi)^2d\sigma }{dxd^2pd^2b}=P(x)
\sum_{k=1}C^n_AT_A^n(b)Y_n(x,p).
\eeq

This iterative procedure can be directly  rewritten  in the $r$ space.
One obtains
\[
Y_n(x,r)=\int \frac{d^2q}{(2\pi)^2}e^{iqr}dwP(w)
I(q,x,w)Y_{n-1}(x_1(q,x,w),r)\]\beq
+\int \frac{d^2q}{(2\pi)^2}e^{iqr}dwP(w)
I(q,x,w)\Big(-\sigma(x_1(q,x,w))\Big)^{n-1}
-\sigma(x)Y_{n-1}(x,r).
\label{eqr}
\eeq

\subsection{The master equation}
Introduce
\beq
Z_n=C_A^nT_A^n(b)Y_n(x,p)
\simeq \frac{1}{n!}(AT_A(b))^nY_n(x,p)\equiv
\frac{1}{n!}\xi^nY_n(x,p),
\eeq
where we denoted $\xi=AT_A(b)$.
Obviously
\beq
\frac{dZ_n}{d\xi}=\frac{1}{(n-1)!}\xi^{n-1}Y_n(x,p).
\eeq

Now  take the recurrent relation (\ref{iter2}) and multiply it by
$\xi^{n-1}/(n-1)!$ to get
\[
\frac{dZ_n}{d\xi}=
\int \frac{d^2q}{(2\pi)^2}dwP(w)
I(q,x,w)\Big\{Z_{n-1}(x_1,p-q)\]\beq
+\frac{1}{(n-1)!}\Big(-\sigma(x_1)\Big)^{n-1}
(2\pi)^2\delta^2(p-q)\Big\}-\sigma(x)Z_{n-1}(x,p).
\eeq
Summing from $n=1$ to $\infty$ and defining
\beq
Z(\xi,x,p)=\sum_{n=1}^{\infty}Z_n,
\eeq
so that
\beq
\frac{(2\pi)^2d\sigma }{dxd^2pd^2b}=P(x)Z(\xi,x,p),
\eeq
we obtain a master equation for $Z$
\[
\frac{dZ(\xi,x,p)}{d\xi}=
\int \frac{d^2q}{(2\pi)^2}dwP(w)
I(q,x,w)\Big\{Z(\xi,x_1,p-q)\]\beq
+e^{-\xi\sigma(x_1)}(2\pi)^2\delta^2(p-q)\Big\}
-\sigma(x)Z(\xi,x,p).
\eeq

Technically this equation does not present serious advantage as
compared to the direct iteration procedure, since it allows to calculate
the contribution only at fixed impact parameter $b$ and requires subsequent
integration over $b$, which for a fixed number of collisions can be done
more easily.

\section{Numerical results}
As mentioned, the iteration procedure allows to reliably calculate
rescattering contributions up to the 16th successive collision.
We choose the gluon mass $m=2$ GeV/c as
an effective infrared cutoff. So our inclusive cross-sections are
meaningful at $p_\perp>2$ GeV/c. We study hA collisions with
gold as a target ($A=197$). For comparison also the copper target was studied
($A=64$). The standard parameter $K$, which effectively
takes into account higher orders in parton scattering cross-sections
has been chosen to be 1.5 at the RHIC energies and 1.0 at the LHC energies.
For the parton distributions we used the CTEQ6L set ~\cite{CTEQ}.

At the RHIC energy ($\sqrt{s}=200$ GeV) the rescattering cross-sections
fall very rapidly with the number $n$ of the subsequent collisions and
are practically zero at $n>6$. In contrast at the LHC energy
($\sqrt{s}=8800$ GeV) the
decrease of the cross-sections with $n$ is much  slower and even
the 12th collision contributes  2 $\div$ 5\% of the total cross-section.

\begin{figure}
\hspace*{1 cm}
\epsfig{file=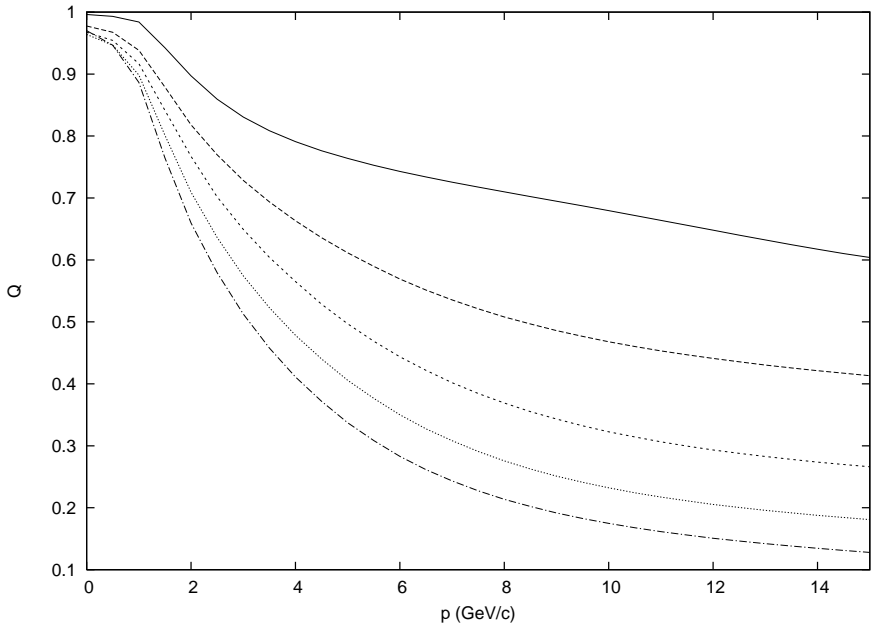,width=12 cm}
\caption{The ratio of the cross-sections with collisional energy loss to the ones without it
(Eq. \ref{qratio}) for p-Au collisions at $\sqrt{s}=200$ GeV.
The curves from top to bottom
correspond to $x=0.025,0.2,0.4,0.6$ and
0.8.}
\label{fig1}
\end{figure}

\begin{figure}
\hspace*{1 cm}
\epsfig{file=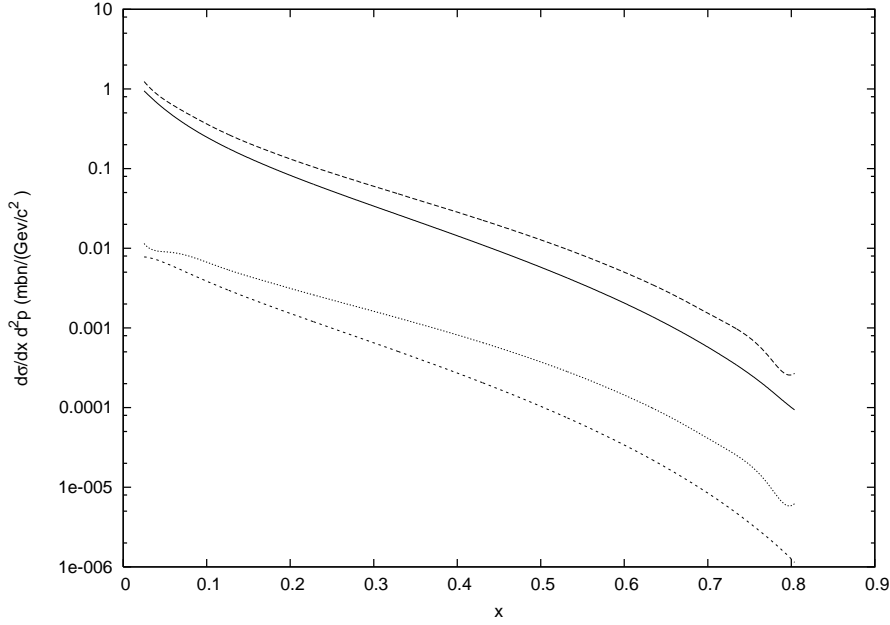,width=12 cm}
\caption{Cross-sections  for p-Au collisions at $\sqrt{s}=200$ GeV
 with and without collisional energy loss as a function of $x$
for $p=5$ GeV/c (upper pair of curves) and $p=10$ GeV/c (lower pair).
In each pair the upper curve corresponds to the absence of energy loss.}
\label{fig2}
\end{figure}

\begin{figure}
\hspace*{1 cm}
\epsfig{file=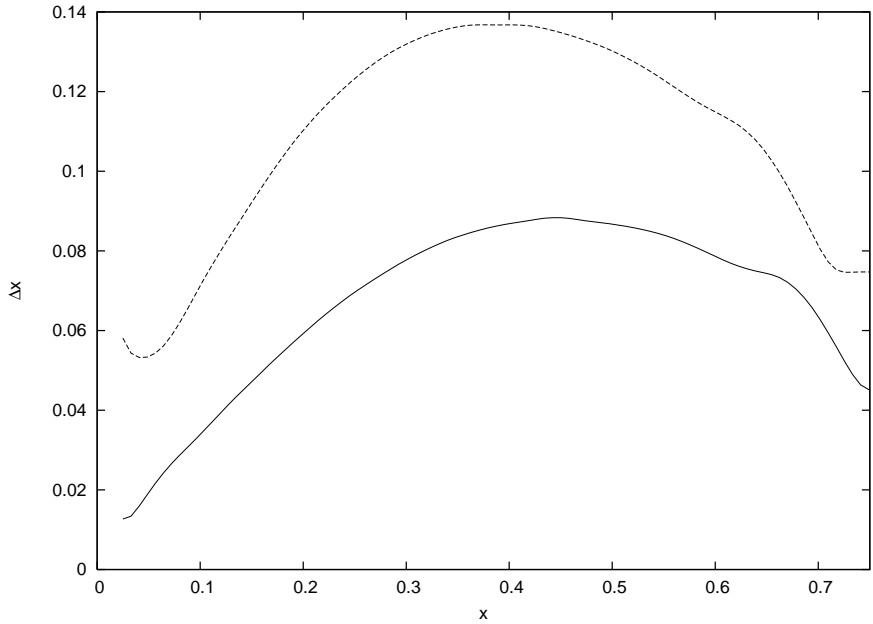,width=12 cm}
\caption{The shift in the scaling variable at $p=5$  GeV/c (lower curve)
and 10 GeV/c  for p-Au collisions at $\sqrt{s}=200$ GeV.}
\label{fig3}
\end{figure}

\begin{figure}
\hspace*{1 cm}
\epsfig{file=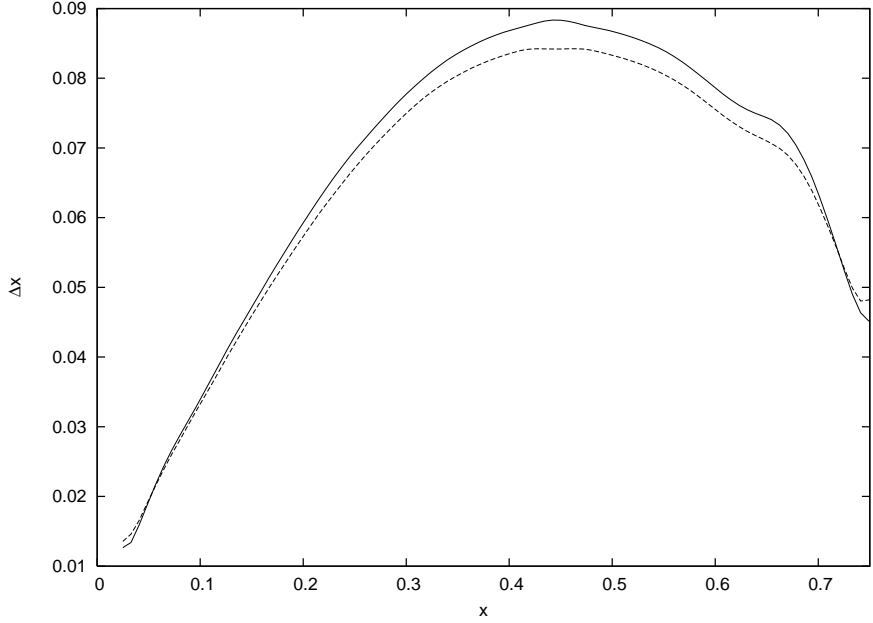,width=12 cm}
\caption{The shift in the scaling variable at $p=5$  GeV/c
for p-Au (the solid curve) and p-Cu  (the dashed curve)
collisions at $\sqrt{s}=200$ GeV.}
\label{fig4}
\end{figure}

\begin{figure}
\hspace*{1 cm}
\epsfig{file=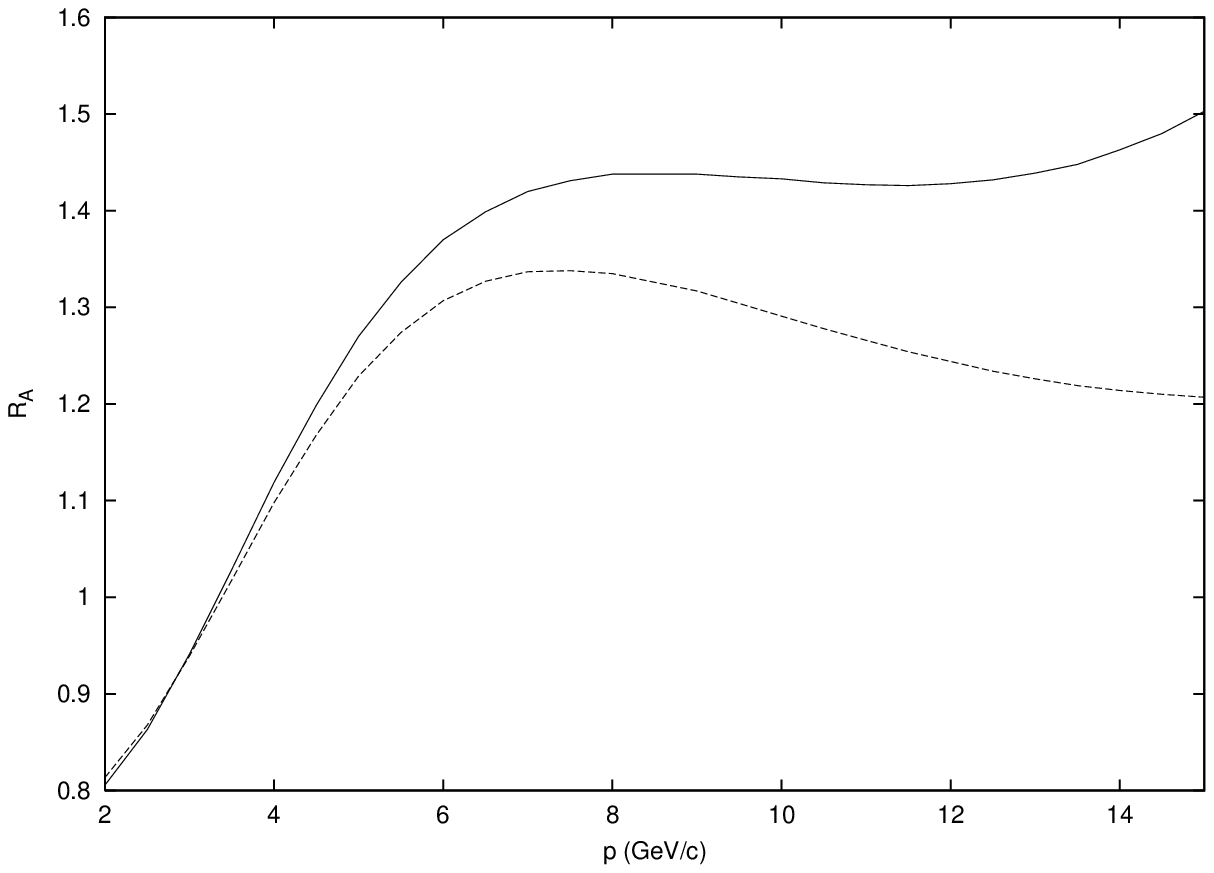,width=12 cm}
\caption{Nuclear suppression factor for p-Au collisions at $\sqrt{s}=
200$ GeV as a function of transverse momentum $p$ at $x=0.025$
with (the solid curve) and without (the dashed curve) collisional
energy loss.}
\label{fig5}
\end{figure}

\begin{figure}
\hspace*{1 cm}
\epsfig{file=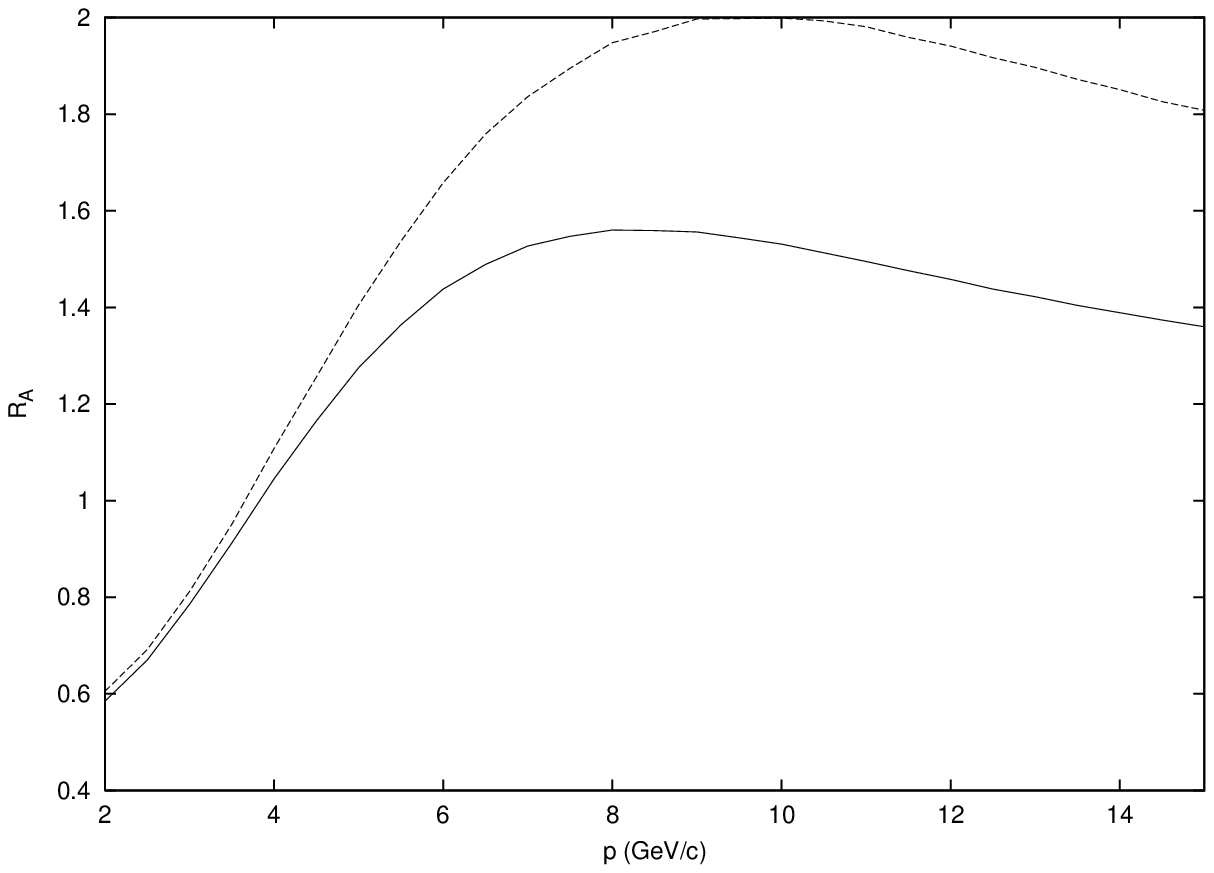,width=12 cm}
\caption{Same as Fig \ref{fig5} at $x=0.4$.}
\label{fig6}
\end{figure}

Our results for the RHIC  energy are presented in Figs.
\ref{fig1}-\ref{fig6}.
In Fig. \ref{fig1} we illustrate the effect of collisional energy loss (CEL)
presenting the ratio of the cross-section with and without CEL
\beq
Q(x,p)=\frac{d\sigma_A}{dx d^2p}\Big/\Big(\frac{d\sigma_A}{dx d^2p}\Big)_{no\ \ energy\ \ loss}
\label{qratio}
\eeq
(In the cross-section with no energy loss $x$ is conserved during rescattering and
is equal to $x$ of the observed jet). As we observe the rate of
CEL grows both with $x$ and $p$, somewhat flattening at $p\geq 10$
GeV/c. In Fig. \ref{fig2} we compare absolute values of the cross-sections with and
without CEL as a function of $x$ at two values of $p=5$ and
10 GeV/c of the transverse momentum. These curves serve us to crudely determine
the magnitude of the energy loss $\Delta$  as a difference between
the values of $x$ at which cross-sections with and without energy loss give the same cross-section
Thus found values of $\Delta$ are presented in Fig. \ref{fig3} as a function
of $x$ \footnote{The chisen primitive procedure to find $\Delta$
fails t high values of $x$ where the cross-sections become
very small, which explains irregularities of the curves at the right end.}.
They show a maximum at  medium values of $x\sim 0.5$ with $\Delta=
0.09$ for $p=5$ and 0.14 for $p=10$. The growth of collision loss with $p$
was to be expected from Eq. (\ref{xxprime}).

An unexpected feature of the CEL is its practical independence of
the atomic number of the target. In Fig. \ref{fig4} we compare the shift
$\Delta$ of the scaling variable for Au and Cu at $p=5$ GeV/c. The two curves
practically coincide. In the common language it implies that the energy loss
does not depend on the length of the path inside the nucleus at all. This can
be understood by the fact that the optimal configuration of rescattering
to give the observed transverse momentum is to produce this momentum
in just one collision, leaving the rest collisions to have a small momentum transfer.
Then obviously it does not matter how many scattering centers there are along
the rescattering path.

Finally in Figs. \ref{fig5} and \ref{fig6} we show the
nuclear suppression factor (NSF)
\beq
R_A(x,p)=\frac{d\sigma_A/(dx d^2p)}{Ad\sigma_1/(dx d^2p)}
\eeq
as a function of $p$  two values of $x=0.025$ and 0.4
with  and without  CEL.
The curious result is  at
low values of $x$ introduction of CEL actually
enhances the NSF
('antiquenching effect'). This is partially the result of the
above mentioned optimal configuration for rescattering in which the effect
of energy loss may be stronger for the single rescattering contribution than for
the total one.

\begin{figure}
\hspace*{1 cm}
\epsfig{file=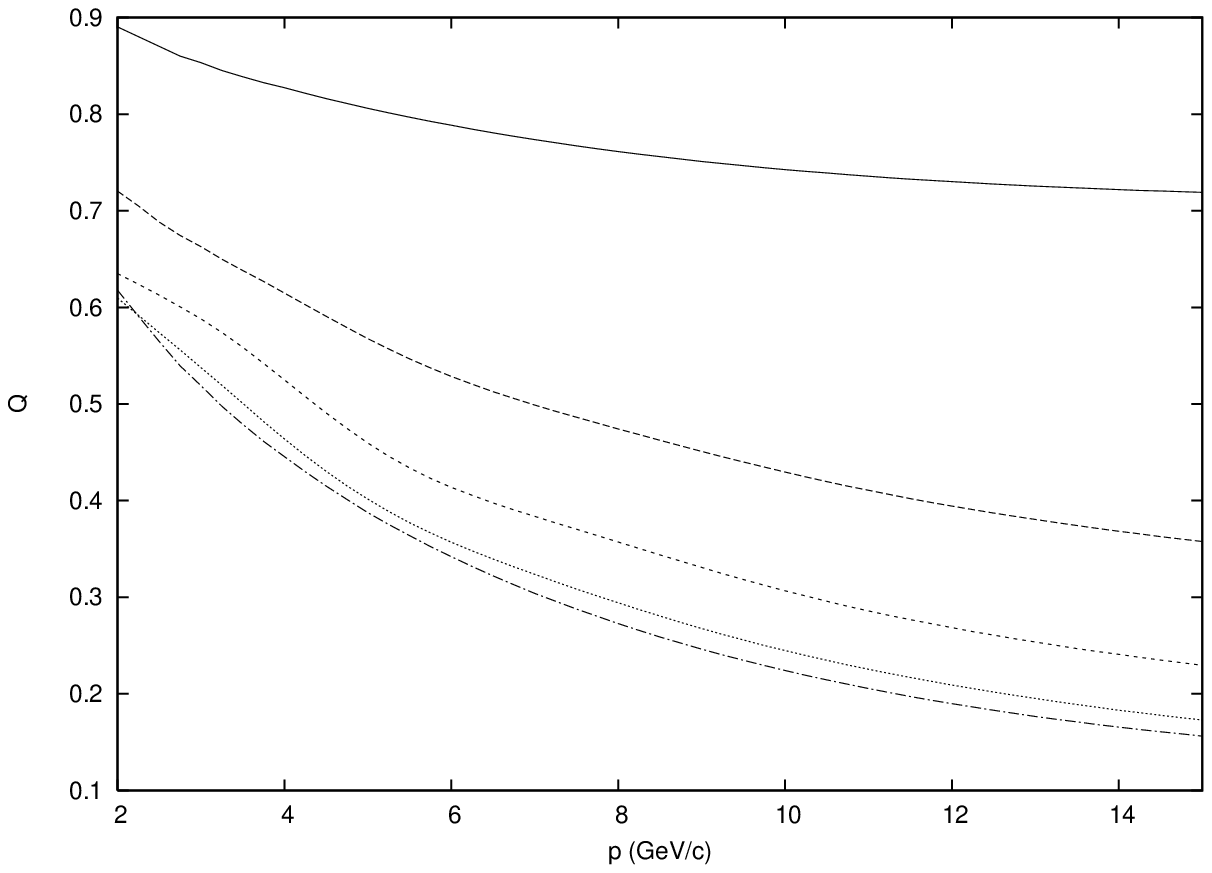,width=12 cm}
\caption{The ratio of the cross-sections with collisional energy loss to the
ones without it (Eq. \ref{qratio}) for p-Au collisions at $\sqrt{s}=8800$ GeV.
The curves from top to bottom
correspond to $x=0.025,0.2,0.4,0.6$ and 0.8.}
\label{fig7}
\end{figure}

\begin{figure}
\hspace*{1 cm}
\epsfig{file=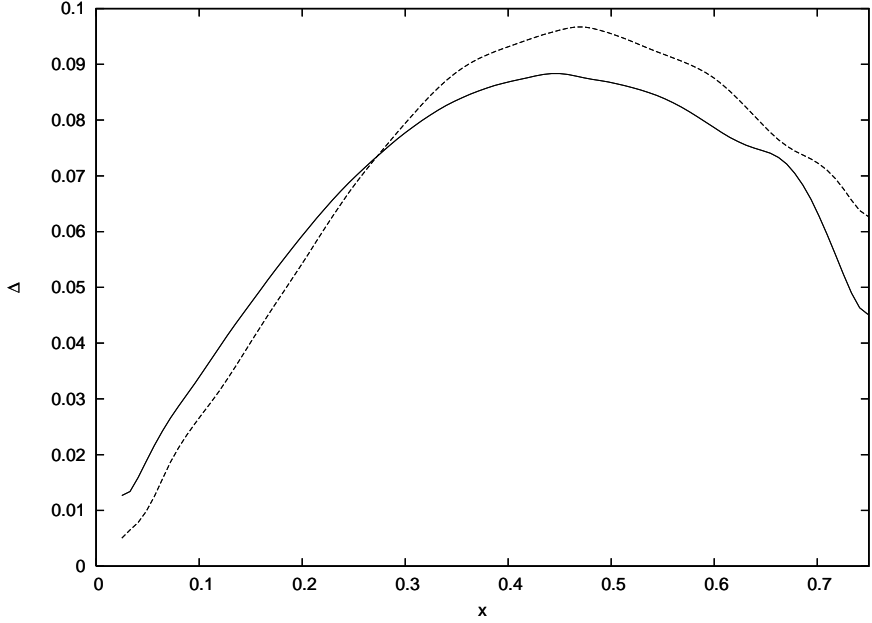,width=12 cm}
\caption{The shift in the scaling variable at $p=5$  GeV/c
for p-Au  collisions at $\sqrt{s}=200$ (the solid curve) and
8800 GeV (the dashed curve).}
\label{fig8}
\end{figure}

\begin{figure}
\hspace*{1 cm}
\epsfig{file=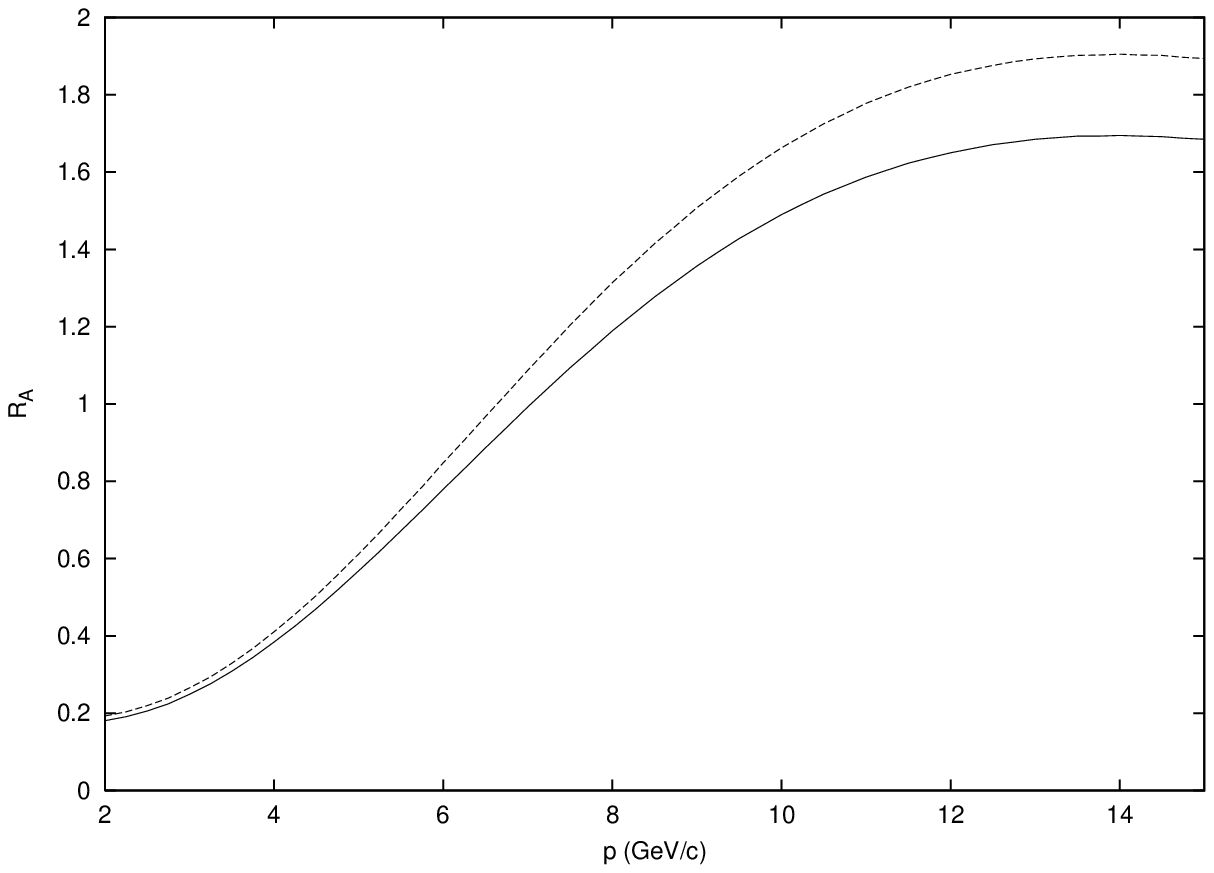,width=12 cm}
\caption{Nuclear suppression factor for p-Au collisions at $\sqrt{s}=
8800$ GeV as a function of transverse momentum $p$ at $x=0.025$
with (the solid curve) and without (the dashed curve) collisional
energy loss}
\label{fig9}
\end{figure}

\begin{figure}
\hspace*{1 cm}
\epsfig{file=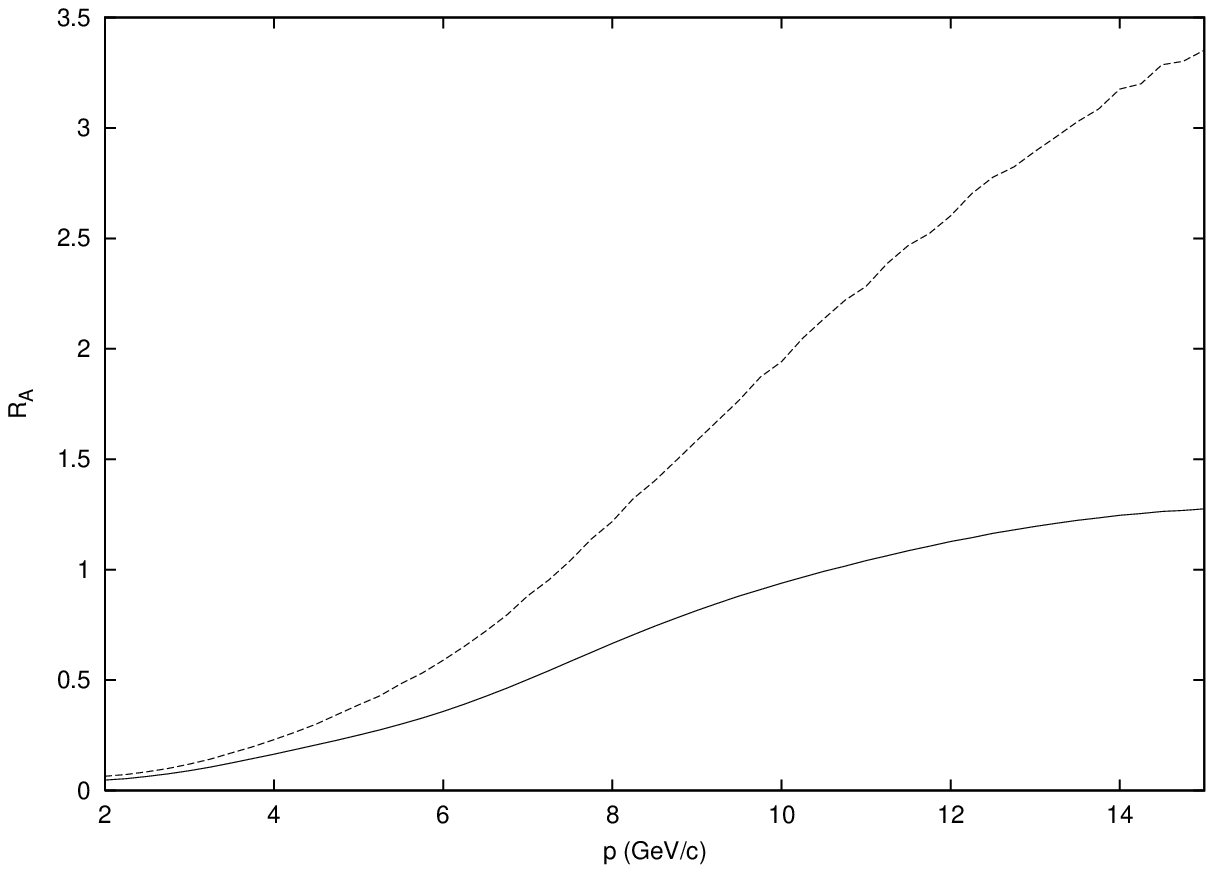,width=12 cm}
\caption{Same as Fig \ref{fig9} at $x=0.4$.}
\label{fig10}
\end{figure}

Our results for the LHC energy $\sqrt{s}=8800$ GeV and Au target
are presented in Figs. \ref{fig7}-
\ref{fig10}. Fig. \ref{fig7} illustrates the CEL showing ratio $Q(x,p)$
(Eq. (\ref{qratio})) as a function of $p$ at different values of $x$.
The general picture remains the same as at the RHIC energy (cf. Fig. \ref{fig1}).
However the $x$ dependence becomes stronger, so that quenching at low $x$ becomes
weaker and at larger $x$ the same or even stronger.
In Fig. \ref{fig8} we compare the shift $\Delta$ of the scaling variable at $p=5$ GeV/c
at LHC and RHIC energies. There is practically no change at all, so that according
to our calculations CEL not only is $A$ independent but also energy
independent. Finally Figs. \ref{fig9} and \ref{fig10} show the
NSF as a function of $p$ at $x=0.025$ and 0.4 respectively. At $x=0.025$ the effect of
CEL is much weaker than at the RHIC energy and acquires the expected
direction: it lowers the NSF. At $x=0.4$ CEL lowers the NSF
much stronger than at the RHIC energy.




\section{Conclusions}
We have calculated collisional energy loss of emitted jets due to multiple collisions
of partons from the projectile hadron with partons in the target nuclei.
Our results indicate that the loss grows with the transverse momentum $p$ of the observed jet
and is maximal at medium values of the scaling variable $x$, where the shift in $x$
reaches values in the interval $0.09-0.14$ for $p$ varying from 5 to 10 GeV/c.
A somewhat unexpected result is that this shift is practically independent on
the atomic number of the target nor on the energy, being the same for RHIC and LHC.
The $A$-independence implies that the shift does not depend on the length passed
by the jet inside the nucleus. An explanation may be that most of the shift takes
place in just one collision during rescattering.

We have also calculated the so-called nuclear suppression factor (NSF) with collisional
energy loss taken into account. From our results it follows that at moderate
transverse momenta NSF still stays noticeably above unity, indicating that energy loss
does not reverse the enhancement in the $p$-distribution due to kicks in
collisions with partons in the target. Surprisingly at the RHIC energy and small $x$
the energy loss even enhances NSF, which again may be the consequence that the loss is mostly
due to just one collision. However one should take into account that our treatment
considers the loss in a single collision  to be relatively small, which condition
may be violated at small values of $xs$ (see Eq. (\ref{xxprime})).

Finally we stress that our results refer to production of jets. To pass to observed
hadrons one should convolute our cross-sections with the appropriate fragmentation functions.
Although this procedure seems direct, we prefer to postpone it for future studies, because
it inevitably involves new badly controlled assumptions and approximations,   especially when
hadronization takes place inside the nucleus. We prefer to present results for jets,
strictly within the perturbational QCD framework, which do not depend on these soft
physics assumptions. They may constitute a starting point for the study of
hadronization process.

\section{Acknowledgments}
The author is most grateful to D.Treleani, who initiated the interest
of the author in this topic. The author is also very indebted to N.Armesto for constructive
discussions, especially concerning literature on the subject.
This work has been supported by grants RFFI 09-012-01327-a and RFFI-CERN.

\end{document}